\documentclass[checkin,showpacs,aps,pra]{revtex4-1}

\usepackage{amsmath,bm}
\newcommand{\br}{{\bf r}}
\newcommand{\kr}{{\bf k, r}}
\newcommand{\hr}{\hat{\bf r}}
\newcommand{\bR}{{\bf R}}

\newcommand{\bk}{{\bf k}}
\newcommand{\bq}{{\bf q}}

\def\bp{\hat{\bf p}}

\newcommand{\hH}{\hat{H}}

\newcommand{\frav}{\frac{1}{V}}

\newcommand{\hh}{\hat{H}}

\newcommand{\ba}{\begin{eqnarray}}
\newcommand{\ea}{\end{eqnarray}}

\usepackage{graphicx,color}

\newcommand{\be}{\begin{equation}}
\newcommand{\ee}{\end{equation}}
\newcommand{\la}{\langle}

\newcommand{\ra}{\rangle}

\newcommand{\hx}{\hat{x}}

\newcommand{\UU}[1]{\label{#1}}

\begin{document}


\title{Revisiting the matrix elements of the position operator in the crystal momentum
  representation}

\author{M. S. Si} \affiliation{School of Materials and Energy, Lanzhou
  University, Lanzhou 730000, China}

\author{G. P. Zhang$^{*}$} \affiliation{Department of Physics, Indiana
  State University, Terre Haute, IN 47809, USA} \date{\today}

\begin{abstract}
  {Fewer operators are more fundamental than the position operator in
    a crystal. But since it is not translationally invariant in
    crystal momentum representation (CMR), how to properly represent
    it is nontrivial.  Over half a century, various methods have been
    proposed, but they often lead to either highly singular
    derivatives or extremely arcane expressions. Here we propose a
    resolution to this problem by directly computing their matrix
    elements between two Bloch states.  We show that the position
    operator is a full matrix in CMR, where the off-diagonal elements
    in crystal momentum $\bk$ only appear along the direction of the
    position vector.  Our formalism, free of singular derivative and
    degeneracy difficulties, can describe an array of physical
    properties, from intraband transitions, polarization with or
    without spin-orbit coupling, orbital angular momentum, to
    susceptibilities.  }
  \end{abstract}

\maketitle

\section{Introduction}
In quantum mechanics, the position operator is a fundamental physical
quantity and has a definite meaning in any nonperiodic system
\cite{dirac,schiff,messiah,landau,griffiths2018,sakurai,ourqm}.  But
it becomes ill-defined in a crystal with a translation symmetry in
crystal momentum representation (CMR)
\cite{kittel,blount1962,sipe2000,marzari2012,bgu2013}. Since the
position operator underlines many physical processes such as optical
excitation \cite{butcher,shen}, polarization
\cite{martin1974,asboth2015,vanderbilt}, and high harmonic generations
\cite{osika2017,parks2023} in solids, a sound description is
necessary.  Over last five decades, at least four methods have been
developed to ameliorate this difficulty.  (i) One redefines it through
an exponential representation of the position operator
\cite{resta1998,aligia1999,evangelisti2022}. This method is often
employed to compute polarization \cite{attaccalite2013}, \UU{change21}
{ and  has only diagonal elements of the position operator in
  crystal momentum $\bk$}.  But for optical transitions, a different
form is used \cite{draxl2006,cazzaniga2010}.  (ii) One converts Bloch
wavefunctions to Wannier wavefunctions \cite{souza2004}, so one works
in the position space instead of in CMR, \UU{change22}{ no
  dependence on $\bk$.}  But this brings in a new difficulty that
Wannier wavefunctions are not eigenstates of the Hamiltonian.  (iii)
One utilizes the commutation relation between the position operator
$\hr$ and Hamiltonian $\hH$ \cite{hughes1996,prb09,esteve2023}, or the
$r$$-$$p$ relation \cite{bgu2013}, $
[\hr,\hH]=\frac{i\hbar}{m_e}\bp. $ One further multiplies both sides
with two Bloch wavefunctions $\psi_n^*(\bk,\br)$ and $\psi_m(\bq,\br)$
from the left and right and integrate to get \be [E_m(\bq)-E_n(\bk)]
\la \psi_n(\bk,\br)|\hr
|\psi_m(\bq,\br)\ra=\frac{i\hbar}{m_e}\la\psi_n(\bk,\br)|\bp|\psi_m(\bq,\br)\ra\delta_{\bk,\bq}, \label{eq1}
\ee where $E_m(\bq)$ and $E_n(\bk)$ are band energies and $m_e$ is the
electron mass.  So one can use the momentum operator to find the
matrix elements of the position operator
\cite{aspnes1972,ghahramani1991,leitsmann2005,allen}, but if two
states are degenerate \cite{sharma2003}, one cannot use it.  (iv) One
takes the partial derivative \cite{karplus1952} of the
orthonormalization between two Bloch wavefunctions with respect to the
crystal momentum \cite{callaway}, $ \frav
\int_{V}\psi_n^*(\bk,\br)\psi_m(\bq,\br)d\br =\delta_{nm}\delta
(\bq-\bk)$, where $V$ is the volume of the crystal. This leads to an
arcane expression \cite{blount1962,lax,matsyshya2019} \be \la
\psi_n(\bk,\br) |\hr |\psi_m(\bq,\br)\ra =-i \frac{\partial \delta
  (\bq-\bk)} {\partial \bq} \delta_{nm}
+\delta(\bq-\bk)\frac{i}{\Omega} \int_\Omega u^*_n(\bk,\br)
\frac{\partial u_m(\bq,\br)}{\partial \bq} d\br, \label{u5} \ee where
the term behind $\delta(\bq-\bk)$ in the second term is the Berry
connection, $\Omega$ is the unit cell volume, and $u_n(\bk,\br)$ is
the cell periodic function.  However, all of the above ignore the
$\bk$-off-diagonal terms of the position operator in CMR, which indeed
exist, as discussed below.

In this paper, we present a resolution by directly computing the
matrix elements between two Bloch wavefunctions, without resorting to
the permutation relation as given in Eq. \ref{eq1} or differentiation
in Eq. \ref{u5}.  We explicitly show that the position operator is
rigorously a full matrix in CMR.  The $\bk$-off-diagonal elements
appear whenever a particular direction loses the translational
symmetry. These elements are the reason for intraband transitions even
in a single band between two $k$ points not adjacent to each other.
In contrast to all the prior studies
\cite{aversa1994,aversa1995,virk2007}, our formalism is even-tempered,
without a singular derivative in Eq. \ref{u5} and free of the
degeneracy problem in Eq. \ref{eq1}, representing a paradigm shift for
future research in a broad scope of fields from nonlinear optics,
ultrafast dynamics, polarization to orbital magnetism and beyond. It
points out a general strategy to implement the position operator in
existing software packages \cite{vasp}, which should be corrected.

\section{Computing matrix elements of the position operator directly}
A straightforward
method \cite{foreman2000} to find the matrix elements of the position
operator between two Bloch wavefunctions $\psi_n(\bk,\br)$ and
$\psi_m(\bq,\br)$ in CMR, without using the derivative above or Berry
connection, would be to directly integrate \be \la \psi_n(\bk,\br) |\hr
|\psi_m(\bq,\br)\ra = \frac{1}{V}\iiint_V \psi_n^*(\bk,\br) \br \psi_m(\bq,\br)
d\br,
\label{main0}
\ee where $d\br$ is the infinitesimal volume element, $\bk$ and $\bq$
are the crystal momenta, and $n$ and $m$ are band indices,
respectively. Both Blount \cite{blount1962} and Foreman
\cite{foreman2000} attempted but failed to proceed through. Instead
they used method (iv), followed by many others
\cite{aversa1994,aversa1995,virk2007}.  When we were searching for a
simple example for our book \cite{ourqm}, we started to revisit this
problem.  Equation \ref{main0} reveals to us immediately that if one
chooses an arbitrary origin at $\br_0$ for $\br$, 
  the $\bk$- and band $n$-diagonal matrix elements depend on $\br_0$,
but the off-diagonal ones, in either $\bk$ or $n$ or both, do
not, because of the orthogonality between two different Bloch
wavefunctions, that is, $\iiint \psi_n^*(\bk,\br) \br_0
\psi_m(\bq,\br) d\br=0$.

The above property gives us a hint. We ought to decompose $\br$ in
Eq. \ref{main0} along three primitive vectors ${\bf a}_1$, ${\bf a}_2$
and ${\bf a}_3$ as shown in Fig. \ref{fig1}(a), so $\br$ can be
written as $\br=\sum r_i\hat{\bf a}_i$, where $r_i$ is the component
along three unit vectors $\hat{\bf a}_i$ with $i$ = 1, 2, and 3.
Then, whenever we shift from the cell by integer multiples of ${\bf
  a}_1$, ${\bf a}_2$, and ${\bf a}_3$, the cell-periodic
$u_n(\bk,\br)$ function remains the same.  Similarly, the volume
element is $d\br=dr_1dr_2dr_3$.  In the following, as an example, we
consider the integral in Eq. \ref{main0} along $\hat{\bf a}_1$, i.e.,
\ba &\frac{1}{V}\int_V\psi^*_n(\bk,\br)r_1 \hat{\bf a}_1
\psi_m(\bq,\br)dr_1 dr_2dr_3 \nonumber \\ =&\frac{\hat{\bf
    a}_1}{V}\int_{r_1}\int_{r_2}\int_{r_3}
u^*_n(\bk,\br)u_m(\bq,\br)e^{
  i(q_1-k_1)r_1+i(q_2-k_2)r_2+i(q_3-k_3)r_3}r_1 dr_1
dr_2dr_3.  \label{u4} \ea It is easy to see that this integral is
cell-periodic along $\hat{\bf a}_2$ and $\hat{\bf a}_3$, leading to
two Kronecker { $\delta_{k_2,q_2}$ and $\delta_{k_3,q_3}$}, but it is
not along $\hat{\bf a}_1$, so $q_1$ may or may not be equal to
$k_1$. In the following, we will consider $q_1\ne k_1$ first
($\bk$-off-diagonal) and then $q_1=k_1$ ($\bk$-diagonal).

\subsection{Segments of the integral}

Before delving into the mathematical details, we briefly outline our
strategy that Blount attempted but failed to proceed
(see the line below Eq. 2.2 on page 309 of Blount's work \cite{blount1962}).
Here, we instead separate
the integral in Eq. \ref{u4}  into three separate integrals over $r_1$, $r_2$ and $r_3$
as $I_1$, $I_2$ and $I_3$, which read
$I_2(r_1,r_3)=\int_{r_2}I_0(\br)e^{i(q_2-k_2)r_2}dr_2$,
$I_3(r_1)=\int_{r_3}I_2(r_1,r_3)e^{i(q_3-k_3)r_3}dr_3$,
and $I_1=\int_{r_1}I_3(r_1)e^{i(q_1-k_1)r_1}r_1dr_1 $ with $I_0(\br)
=u^*_n(\bk,\br)u_m(\bq,\br)$.  $I_2$ and $I_3$ inherit the lattice
periodicity of $I_0(\br)$, but $I_1$ does not because of $r_1$.
However, $I_1$, $I_2$ and $I_3$ share a similar structure and can be
written as a position-dependent phase factor multiplied by a common
integral. We integrate it in the order of $I_{2} \rightarrow I_{3} \rightarrow I_{1}$.

In the following, we illustrate the integral of $I_2$ first.
We need to break $I_2(r_1,r_3)=\int_{r_2=0}^{N_2a_2} I_0(\br)e^{i(q_2-k_2)r_2}
dr_2$ into $N_2$ segments along the $\hat{\bf a}_2$ axis
\cite{esteve2023} (see Fig. \ref{fig1}(a)) as $ I_2(r_1,r_3)=
(\int_{r_2=0}^{a_2}+\int_{r_2=a_2}^{2a_2}+\cdots
+\int_{r_2=(N_2-1)a_2}^{N_2a_2}) I_0(\br)e^{i(q_2-k_2)r_2} dr_2.  $ We
denote the first integral as  $ I_{2,1}
=\int_{r_2=0}^{a_2}I_0(\br)e^{i(q_2-k_2)r_2} dr_2.$ Since $I_0(\br)$
is periodic by an integer shift along ${\bf a}_2$, the second integral
$ I_{2,2}=\int_{r_2=a_2}^{2a_2}I_0(\br)e^{i(q_2-k_2)r_2} dr_2 $ can be
simplified by introducing a new variable $\xi_2=r_2-a_2$, to get  $
I_{2,2}=\int_{\xi_2=0}^{a_2}I_0(\br)e^{i(q_2-k_2)(\xi_2+a_2)}
d\xi_2\nonumber
=e^{i(q_2-k_2)a_2}\int_{\xi_2=0}^{a_2}I_0(\br)e^{i(q_2-k_2)\xi_2}
d\xi_2= e^{i(q_2-k_2)a_2} I_{2,1}, $ which differs from
$I_{2,1}$ by a factor $e^{i(q_2-k_2)a_2}$. This is the key
to integrate $I_{2}$ analytically.

Then the remaining terms, such as  
the last integral $I_{2,N_2}$, can be calculated similarly,
$I_{2,N_2}= e^{i(q_2-k_2)(N_2-1)a_2} I_{2,1}$, 
so the entire integral $I_2$ is just a geometric series, 
\be
I_2(r_1,r_3)=(1+e^{i(q_2-k_2)a_2}+\cdots+e^{i(N_2-1)(q_2-k_2)a_2})I_{2,1}=
I_{2,1}
\frac{1-e^{iN_2(q_2-k_2)a_2}}{1-e^{i(q_2-k_2)a_2}}=I_{2,1}N_2\delta_{k_2,q_2},
\label{supu5} 
\ee
where we have used the Born-von Karman (BvK) boundary condition, 
$\delta_{k_2,q_2}$ is a Kronecker delta and $N_2$ is the number of
cells along ${\bf a}_2$.
Similarly the integration over $r_3$ is 
$I_3(r_1)=I_{3,1}N_3\delta_{k_3,q_3}$,
where $ I_{3,1}=
\int_{r_3=0}^{a_3}
I_{2}(r_1,r_3) e^{i(q_3-k_3)r_3}
dr_3.$

\subsection{The integral of $I_{1}$}

The most complicated integral is over $r_1$, but we can carry out a
similar calculation. We also split the integral into $N_1$ segments as
\be I_1=\int_{r_1=0}^{N_1a_1} I_3(r_1)e^{i(q_1-k_1)r_1}r_1dr_1 =\left
(\int_{r_1=0}^{a_1}+\int_{a_1}^{2a_1}+\cdots+\int_{(N_1-1)a_1}^{N_1a_1}\right)
I_3(r_1)e^{i(q_1-k_1)r_1}r_1dr_1. \label{u6} \ee We denote the first
integral as \be I_{1,1}=\int_{r_1=0}^{a_1}
I_3(r_1)e^{i(q_1-k_1)r_1}r_1dr_1.  \ee The second integral is \be
I_{1,2}=\int_{a_1}^{2a_1}
I_3(r_1)e^{i(q_1-k_1)r_1}r_1dr_1, \label{u61} \ee which can be
simplified as above by setting $r_1-a_1=\xi_1$ to obtain, \ba
I_{1,2}&=&\int_{0}^{a_1}
I_3(r_1)e^{i(q_1-k_1)(\xi_1+a_1)}(\xi_1+a_1)d\xi_1\\ &=&e^{i(q_1-k_1)a_1}
\left (\int_{0}^{a_1} I_3(r_1)e^{i(q_1-k_1)\xi_1}\xi_1
d\xi_1+a_1\int_{0}^{a_1} I_3(r_1)e^{i(q_1-k_1)\xi_1}
d\xi_1\right)\\ &=&e^{i(q_1-k_1)a_1}(I_{1,1}+a_1 Q), \label{u62} \ea
where the last equation serves the definition of $Q$. Next, we work
out the last term, \be I_{1,N_1}=\int_{(N_1-1)a_1}^{N_1a_1}
I_3(r_1)e^{i(q_1-k_1)r_1}r_1dr_1, \ee by setting
$r_1-(N_1-1)a_1=\xi_1$.  So our $ I_{1,N_1}$ becomes \ba I_{1,N_1}&=&
\int_0^{a_1}
I_3(r_1)e^{i(q_1-k_1)(\xi_1+(N_1-1)a_1)}(\xi_1+(N_1-1)a_1)d\xi_1\\ &=&e^{i(q_1-k_1)(N_1-1)a_1}
\left (\int_{0}^{a_1} I_3(r_1)e^{i(q_1-k_1)\xi_1}\xi_1
d\xi_1+(N_1-1)a_1\int_{0}^{a_1} I_3(r_1)e^{i(q_1-k_1)\xi_1}
d\xi_1\right)\nonumber\\ &=&e^{i(q_1-k_1)(N_1-1)a_1}(I_{1,1}+(N_1-1)a_1
Q).  \ea Then, our $I_1$ is an elegant summation over two series \ba
I_1&=&I_{1,1}\left
(1+e^{i(q_1-k_1)a_1}+\cdots+e^{i(q_1-k_1)(N_1-1)a_1}\right)\nonumber\\ &+&Q\left
(a_1 e^{i(q_1-k_1)a_1}+\cdots+(N_1-1)a_1 e^{i(N_1-1)(q_1-k_1)a_1}
\right), \label{I1} \ea whose first series is the same one in
Eq. \ref{supu5} and the second one can be found by taking the
derivative of the first one with respect to $(q_1-k_1)$, then divided
by $i$, i.e., \be \frac{\partial
  (1+e^{i(q_1-k_1)a_1}+\cdots+e^{i(q_1-k_1)(N_1-1)a_1})} {i\partial
  (q_1-k_1)}=a_1 e^{i(q_1-k_1)a_1}+\cdots+(N_1-1)a_1
e^{i(N_1-1)(q_1-k_1)a_1}.  \ee Since \be
1+e^{i(q_1-k_1)a_1}+\cdots+e^{i(q_1-k_1)(N_1-1)a_1}=
\frac{1-e^{iN_1(q_1-k_1)a_1}}{1-e^{i(q_1-k_1)a_1}}, \ee its derivative
is \be \frac{-iN_1a_1e^{iN_1(q_1-k_1)a_1}}{1-e^{i(q_1-k_1)a_1}}
+\frac{(1-e^{iN_1(q_1-k_1)a_1})(ia_1e^{i(q_1-k_1)a_1})}
{(1-e^{i(q_1-k_1)a_1})^2}. \label{deriv1} \ee Using the BvK boundary
condition, $e^{iN_1(q_1-k_1)a_1}=1$, the second term is eliminated,
while the first term is $\frac{-iN_1a_1}{1-e^{i(q_1-k_1)a_1}}$ 
  with $q_1\ne k_1$, denoted as $(1-\delta_{k_1,q_1})$
  below. See more discussions in \cite{sm}.

Then  Eq. \ref{I1} is 
\be
I_1=I_{1,1}N_1\delta_{k_1,q_1}+Q\frac{-N_1a_1}{1-e^{i(q_1-k_1)a_1}},
\label{main1}
\ee
where the  first term is zero because of  $q_1\ne k_1$, and 
the second term, after writing out all the terms in $Q$,
is
\be
\frac{-a_1
N_1N_2N_3\delta_{k_2,q_2}\delta_{k_3,q_3}
}{1-e^{i(q_1-k_1)a_1}}
\int_{r_1=0}^{a_1}\int_{r_2=0}^{a_2}\int_{r_3=0}^{a_3}
u^*_n(\bk,\br)u_m(\bq,\br)e^{i(q_1-k_1)r_1} e^{i(q_2-k_2)r_2}
e^{i(q_3-k_3)r_3}dr_1 dr_2dr_3. \nonumber
\ee
{ It can be  written in the shorthand notation as}
\be
N_1N_2N_3\delta_{k_2,q_2}\delta_{k_3,q_3}\frac{-a_1}{1-e^{i(q_1-k_1)a_1}}
\int_{\Omega}
u^*_n(\bk,\br)u_m(\bq,\br)e^{i(\bq-\bk)\cdot \br}
d\br. \label{eq3}
\ee

Finally, we consider the $\bk$-diagonal term
  ($k_j=q_j$) in Eq. \ref{u4}, where the exponentials become 1. Since
  $u_n^*$ and $u_m$ are cell-periodic, the integral over $r_2(r_3)$ is
  just $N_2(N_3)$ times the integral over the primitive cell. The
  integral over $r_1$ follows the same steps as in Eq. \ref{u6}. A
  straightforward calculation \cite{sm} gives $ \frac{\hat{\bf
      a}_1}{V}\int_{r_1}\int_{r_2}\int_{r_3}
  u^*_n(\bk,\br)u_m(\bq,\br)r_1 dr_1 dr_2dr_3
  =\frac{N_1N_2N_3}{V}\delta_{\bq,\bk} \left (\int_\Omega
  u^*_n(\bk,\br)u_m(\bq,\br){\bf r}_1 dr_1 dr_2dr_3+\frac{(N_1-1){\bf
      a}_1}{2}\int_\Omega u^*_n(\bk,\br)u_m(\bq,\br)
  dr_1dr_2dr_3\right ),$ where ${\br_1}$ must be understood as a
  component along ${\bf a}_1$ direction.   Since $V=N_1N_2N_3\Omega$,
the matrix elements of the position operator along the ${\bf a}_1$
direction in Eq. \ref{u4}, $\frac{1}{V}\int_V\psi^*_n(\bk,\br)\br_1
\psi_m(\bq,\br)d\br$, can be written as {a sum of
  $\bk$-off-diagonal and   $\bk$-diagonal 
integrals over the primitive
cell,} {\ba &&(1-\delta_{k_1,q_1})
  \delta_{k_2,q_2}\delta_{k_3,q_3}
  \frac{-a_1}{1-e^{i(q_1-k_1)a_1}}\frac{\hat{\bf a}_1}{\Omega}
  \int_\Omega u^*_n(\bk,\br)u_m(\bq,\br) e^{i(\bq-\bk)\cdot \br} d\br
  \nonumber \\ &+&\delta_{\bk,\bq}\frac{1}{\Omega} \left ( \int_\Omega
  u^*_n(\bk,\br) {\br_1} u_m(\bq,\br) d\br+ \frac{(N_1-1){\bf
      a}_1}{2}\int_\Omega u^*_n(\bk,\br)u_m(\bq,\br)
  d\br\right). \label{u64} \ea }

\subsection{The complete integral of the matrix elements of the position operator $\hat{\bf r}$}

\begin{figure}
  \includegraphics[angle=0,width=0.75\columnwidth]{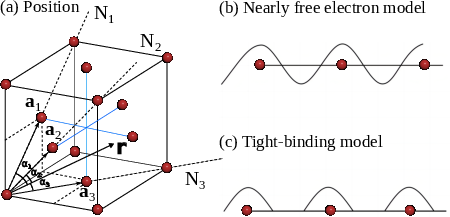}
  \caption{ (a) Lattice points for the integration in Eq. \ref{main0}. The
  crystal has dimension of $N_1{\bf a}_1\times N_2{\bf a}_2\times
  N_3{\bf a}_3$, where $\{{\bf a}_i\}$ are three lattice vectors.
{ $\alpha_i$ represents the angle of ${\bf r}$ with respect to
    ${\bf a}_{i}$.}
The Born-von Karman periodic boundary conditions are
  imposed along three axes. (b) Nearly free-electron model. (c)
  Tight-binding model. }
\label{fig1}
\end{figure}

Along other two directions, we can use the same method.  For an
arbitrary position vector $\br$ that makes an angle of $\alpha_i$ with
respect to the ${\bf a}_i$ axis (see Fig. \ref{fig1}(a)), its matrix
elements contain $\bk$-off-diagonal and $\bk$-diagonal terms, \ba
&\la\psi_n(\bk,\br) |\hr |\psi_m(\bq,\br)\ra \nonumber
\\ =&\frac{\bm{{\cal R}}}{\Omega}\int_\Omega
u^*_n(\bk,\br)u_m(\bq,\br) e^{i(\bq-\bk)\cdot \br}d\br +
\frac{\delta_{\bk,\bq}}{\Omega} \int_\Omega
u^*_n(\bk,\br)(\br+\bm{{\cal T}}) u_m(\bq,\br)d\br, \label{main} \ea
where $\bm{{\cal R}} =-\frac{(1-\delta_{q_1,k_1}) \delta_{q_2,k_2}
  \delta_{q_3,k_3}\cos\alpha_1}{1-e^{i(q_1-k_1)a_1}} {\bf a}_1
-\frac{\delta_{q_1,k_1}(1-\delta_{q_2,k_2})
  \delta_{q_3,k_3}\cos\alpha_2 }{1-e^{i(q_2-k_2)a_2}} {\bf a}_2
-\frac{\delta_{q_1,k_1}\delta_{q_2,k_2}(1-\delta_{q_3,k_3})\cos\alpha_3
}{1-e^{i(q_3-k_3)a_3}} {\bf a}_3$, $(1-\delta_{k_i,q_i})$ excludes
$k_i=q_i$, yielding the $\bk$-off-diagonal elements, $k_i$ and $q_i$
are three components of $\bk$ and $\bq$ along three reciprocal lattice
vectors ${\bf b}_1,{\bf b}_2,$ and ${\bf b}_3$, respectively, and
$\cos\alpha_i$ is the direction cosine.  \UU{change5} { Because
  the crystal structure consists of the lattice points and the basis
  (unit cells) \cite{kittel}, the lattice manifests itself in the
  matrix elements of the position operator through $\bm{{\cal
      T}}=\frac{N_1-1}{2}{\bf a}_1+\frac{N_2-1}{2}{\bf
    a}_2+\frac{N_3-1}{2}{\bf a}_3 $, a vector at the center of the
  crystal, extending the Zak's band center position \cite{zak1982} to
  the entire crystal. If the entire crystal has one cell,
  $N_1=N_2=N_3=1$, $\bm{{\cal T}}=0$, nicely recovering the molecule
  limit.}

Equation \ref{main} is our main result, and has some unfamiliar
properties that are not shared by prior formulations.  First, in
contrast to Eqs. \ref{eq1} and \ref{u5}, our equation has neither a
degenerate problem \cite{blount1962,foreman2000} nor a singular
derivative of the Dirac $\delta$ function
\cite{blount1962,aversa1995}, a huge advantage over the prior schemes
\cite{blount1962,aversa1995}. Second, both integrals are over the
primitive unit cells, {\it not over the entire space}.  The first
integral will never produce a Kronecker $\delta$ in $\bk$ as the Bloch
functions are not orthogonal in general in the primitive cell
\cite{shirley1996,prendergast2009} and the orthogonality only appears
when the integral is over the entire space (see \cite{callaway}).
Third, whenever the matrix element is diagonal in both $\bk$ and band
index $n$, \UU{change3} { there are two major contributions
  because of the crystal structure discussed above.  One comes from
  the lattice,  $\bm {{\cal T}}/\Omega$ (see SM for
  details \cite{sm}), so the electric 
polarization is given as ${\bf P} = -e\bm
  {{\cal T}}/\Omega$ as seen before \cite{vanderbilt}.  The other is
  the integral over {\bf r} within the unit cell, 
$\frac{1}{\Omega}\int_\Omega u^*_n(\bk,\br) \br u_m(\bq,\br)d\br$.}
This integral depends on the origin of the
coordinates. Suppose we shift $\br$ to $\br+\br_0$, where $\br_0$ is
less than $\{ {\bf a}_i\}$. Then the cell-periodic $u_n(\kr+\br_0)$,
which is periodic only in terms of multiples of ${\bf a}_i$, can be
quite different from $u_n(\kr)$, so is the integral, besides a rigid
shift by $\br_0$. This may explain why choosing a different unit cell
produces a different polarization \cite{vanderbilt}\UU{change1} {
  (see an example in SM \cite{sm}).  The ambiguity in both the crystal
  center and the second integral can be traced back to the
  arbitrariness of the coordinate origin in Eq. \ref{main0}.}  But as
soon as $n\ne m$ or $\bk\ne\bq$, this arbitrariness is
removed. Therefore, when we compute the polarization change, the
result is unique since it involves changes in $\bk$ or $n$ or both.

 Fourth, $\bm{{\cal R}}$ further reveals that the $\bk$-off-diagonal
 term only appears along the direction of $\br$.  For instance, if in
 the real space, $\br$ is along ${\bf a}_1$, then in the $\bk$ space,
 $q_1\ne k_1$. The factor $\frac{1}{1-e^{i(q_1-k_1)a_1}}$ in
 $\bm{{\cal R}}$ is produced when $\br$ is beyond the primitive
 cell. Take Eq. \ref{u61} as an example, whose integration limits of
 $I_{1,2}$ are $[a_1,2a_1]$, falling outside the primitive cell
 $[0,a_1]$.  Once we fold it back to the primitive cell, it contains
 $Q$ in it (see Eq. \ref{u62}). This ultimately produces Eq. \ref{eq3}
 and finally the first term in Eq. \ref{main}. If $\br$ points in an
 arbitrary direction (Fig. \ref{fig1}(a)), then all the three possible
 $\bk$-off-diagonal terms $(q_i\ne k_i)$ appear, weighted by their
 respective direction cosines $\cos\alpha_i$.  Physically, the
 presence of $\br$ destroys the translational symmetry and introduces
 a phase factor. This allows an electric field to move electrons even
 within a single band from one $\bk$ to another, and even when two
 $\bk$ points are not adjacent to each other, regardless of how small
 the field is, as far as the Pauli exclusion principle allows.  This
 represents a paradigm shift.

Our even-tempered form greatly eases numerical implementations,
without transforming Bloch wavefunctions to Wannier wavefunctions or
the Berry connection as done currently \cite{vanderbilt}.  For
instance, we can compute the matrix elements of the orbital angular
momentum $\hat{\bf L}$ between $|n_1\bk_1\ra$ and $|n_2\bk_2\ra$, \be
\la n_1\bk_1|\hat{\bf L}|n_2\bk_2\ra=\sum_m\la
n_1\bk_1|\hat{\br}|m\bk_2\ra \times \la m\bk_2|{\bp}|n_2\bk_2\ra, \ee
which has $\bk$-off-diagonal elements as well.  With the spin-orbit
coupling, our method has a clear advantage over
Eq. \ref{eq1}. Consider the Bloch wavefunction
$|\psi_{n\bk}\ra=a_{n\bk}|{\uparrow}\ra +b_{n\bk}|{\downarrow}\ra $,
where $a_{n\bk}$ and $b_{n\bk}$ are the spin majority $|{\uparrow}\ra$
and minority $|{\downarrow}\ra$ parts of the wavefunction. Then the
matrix element of the position operator between two Bloch states
$|\psi_{n\bk}\ra$ and $|\psi_{m\bq}\ra$ is \be \la
\psi_{n\bk}|\hr|\psi_{m\bq}\ra= \la a_{n\bk}|\hr| a_{m\bq}\ra + \la
b_{n\bk}|\hr| b_{m\bq}\ra.  \ee

\section{Examples}

In this section, we present two analytical examples to validate our
formula given above.  One is the nearly free-electron model and the
other is the tight-binding model. The former has highly delocalized
electrons while the latter binds the electrons tightly to the
nuclei. They represent two limiting cases, whereas all the other
models fall in between them.

\subsection{A nearly free-electron model }

Figure \ref{fig1}(b)
shows the nearly free-electron model in a one-dimensional chain of
length $L=Na$ with a periodic potential $\hat{U}(x)$, where the
crystal Hamiltonian is $ \hh=-\frac{\hbar^2}{2m_e}\frac{d^2}{d
  x^2}+\hat{U}(x). $ The wavefunction $\psi(k,x)$ \cite{sm}, up to the first
order, is \be \psi(k,x)=\frac{{\cal C}}
{\sqrt{L}}\left(1+\sum_{n\ne0}\frac{U_{n}e^{inbx}}{
  \frac{\hbar^{2}k^{2}}{2m_e}-\frac{\hbar^{2}(k+nb)^{2}}{2m_e}
          }
          \right) e^{ikx}
\equiv \frac{{\cal C}}{\sqrt{L}}\left (1+\sum_{n\ne0} W_n(k) e^{inbx}\right)e^{ikx},
\label{ch6.eq1212b}
\ee where ${\cal C}$ is the normalization constant,
$b=\frac{2\pi}{a}$, $U_n$ is the Fourier coefficient of $\hat{U}(x)$,
and the dimensionless $W_n$ is defined through Eq. \ref{ch6.eq1212b}.
Since $\psi(k,x)$ is a single band, we do not include a band
index. The matrix element $\la \psi(k_1,x)|\hx|\psi(k_2,x)\ra$ for
$k_1\ne k_2$ is \cite{sm}
\be \frac{|{\cal C}|^2}{i(k_2-k_1)}
+\sum_{m\ne0}\frac{|{\cal C}|^2W_m(k_2)}{i(mb+k_2-k_1)}
+\sum_{n\ne0}\frac{|{\cal
    C}|^2W_n^*(k_1)}{i(k_2-nb-k_1)}+\sum_{n\ne0,m\ne0}\frac{|{\cal
    C}|^2W_n^*(k_1) W_m(k_2)}{i(mb+k_2-nb-k_1)}, \label{ch6.eq1212c}
\ee which explicitly
shows the position operator is not diagonal in $k$ and is not
singular.

\subsection{Tight-binding model}

\begin{figure}
  \includegraphics[angle=0,width=0.8\columnwidth]{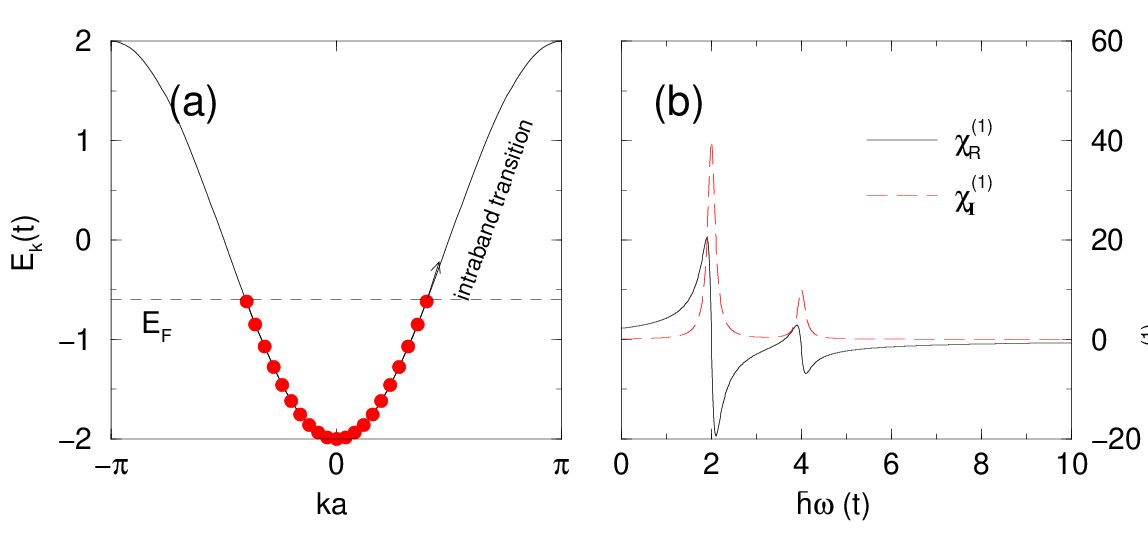}
  \caption{ (a) Band dispersion, with the Fermi level denoted by the
    dashed line $E_F$. The filled circles are occupied states.  The
    arrow shows that an external electric field only can induce
    intraband transitions.  (b) Real (solid line) and imaginary
    (dashed line) susceptibilities as a function of the photon energy
    $\hbar\omega$ in units of hopping integral $t$. Two peaks come
    from two energy gaps.  }
\label{fig2}
\end{figure}

Figure \ref{fig1}(c) shows the
one-dimensional and single-band tight-binding model. Choosing a single
band is purposeful. If the position operator had no off-diagonal
elements in $k$, then there would be no intraband transition and 
the system would not respond to an electric
field. 
Thus, this model represents a
litmus test for any theory.  Our Hamiltonian is \cite{ourqm},
$\hH=\sum_{s}t_{0}\vert s\rangle \langle s \vert-t\sum_{s}
(\vert s+1\rangle \langle s \vert+\vert s\rangle \langle s+1 \vert),$
where $s$ is the site index and runs from $1$ to $N$,  $\vert s\rangle$
is the orthonormalized atomic orbital, and $\la
s_1|s_2\ra=\delta_{s_1,s_2}$. 
Its eigenenergy is  $E(k)=t_0-2t\cos(ka)$ and 
its eigenstate is 
$\psi(k)=\frac{1}{\sqrt{N}}
\sum_{s}e^{iska}|s\rangle.$
Figure \ref{fig2}(a) shows the band structure with the Fermi level
denoted by  $E_F$. 
  The
position operator is 
$\hat{x}=\sum_{s} sa |s \rangle \langle s|.$
Its diagonal element, i.e., its expectation value, is \be \langle
\psi(k) |\hat{x}| \psi(k) \rangle=\frac{1}{N}
\sum_{s,s_1,s_2}e^{-is_1ka}e^{is_2ka}sa\la s_1|s\ra\la s|s_2\ra
=\frac{1}{N} \sum_{s}sa=\frac{(N-1)a}{2},\label{matrix0} \ee which is
in the middle of the chain as expected.  The expectation value of
polarization $\hat{P}=-e\hx/L$ is finite even at $N\rightarrow
\infty$, $ P_0=-\frac{e\la\psi(k)|\hx |\psi(k)\ra
}{L}=-\frac{e}{2}(1-\frac{1}{N}) \xrightarrow{N\rightarrow \infty}
-\frac{e}{2}$ \cite{zak1982,zak1989}. Its off-diagonal
element is 
$ \la \psi(k_1)|\hat{x}|\psi(k_2)\ra=\frac{1}{N}\sum_s
e^{is(k_2-k_1)a}sa= -\frac{a}{1-e^{i(k_2-k_1)a}}.$ To demonstrate that
the position operator really forms a matrix, we choose $N=4$
and { use the BvK boundary condition} to get \be
\la\psi(k_1)|\hat{x}|\psi(k_2)\ra= \frac{a}{2}
 \begin{pmatrix}
   3& -\gamma & -1 &-\gamma^* 
   \\
   -\gamma^* & 3 &-\gamma& -1 \\
   -1 & -\gamma^*& 3 & -\gamma\\
   -\gamma & -1 & -\gamma^*& 3\\
 \end{pmatrix},
 \label{matrix}
 \ee where $\gamma=1+i$. This is a $4\times 4$ matrix. 

Now we come to our most stringent litmus test.  Suppose that initially
the system is in an unperturbed state $|\psi(k_1)\ra$.  Under a small
electric field $F$ along the $x$ axis,
{ the  perturbative
wavefunction up to the first order is
$|\tilde{\psi}(k_1)\ra={\cal C}\left (|\psi(k_1)\ra+\sum_{k_2\ne k_1} \frac{eF\la
  \psi(k_2) |\hx|\psi(k_1)\ra}{E(k_1)-E(k_2)}|\psi(k_2)\ra\right)$, where
 ${\cal C}$ is the normalization constant up to the first order}.  The
polarization is given by $ \la \tilde{\psi}(k_1)|\hat{P}
|\tilde{\psi}(k_1)\ra=P_0+P_1+P_2 =P_0+\epsilon_0 \chi^{(1)}F
+\epsilon_0 \chi^{(2)}F^2$, where the first-order ($\chi^{(1)}$) and
second-order ($\chi^{(2)}$) susceptibilities \cite{shen,butcher,prb00}
are \ba \chi^{(1)}&=&2\frac{e^2{\cal C}^2}{\epsilon_0L}\sum_{k_2\ne k_1}
\frac{|\la \psi(k_1)|\hx|\psi(k_2)\ra|^2}{E(k_2)-E(k_1)}, \label{x1}
\\ \chi^{(2)}&=&\frac{(-e)^3{\cal C}^2}{\epsilon_0 L} \sum_{k_2\ne k_1,k_3\ne
  k_1} \frac{ \la\psi(k_1)|\hx|\psi(k_3)\ra
  \la\psi(k_3)|\hx|\psi(k_2)\ra
  \la\psi(k_2)|\hx|\psi(k_1)\ra}{(E(k_1)-E(k_2))
  (E(k_1)-E(k_3)) \label{xi2} }. \label{x2} \ea If the position
operator were diagonal in $k$ as assumed previously
\cite{aversa1994,aversa1995,virk2007,parks2023}, $\chi^{(1)}$ and
$\chi^{(2)}$ would be zero immediately since we only have a single
band (see the arrow in Fig. \ref{fig2}(a)).  This is clearly
unphysical.  Our example is the manifestation of our new paradigm
above.  It is also true for optical excitation.  We take our above
$N=4$ as an example, where $k_n= \frac{n}{N}\frac{2\pi}{a}$ and
$E(k_1)=t_0$, $E(k_2)=t_0+2t$, $E(k_3)=t_0$ and
$E(k_4)=t_0-2t$. Assume initially $E(k_4)$ is singly occupied. We can
directly compute the susceptibility. Figure \ref{fig2}(b) shows the
real and imaginary susceptibilities as a function of photon energy
$\hbar\omega$. Two peaks come from two different energy gaps. The
excitation is due to the intraband transition.  These peaks solely
originate from these off-diagonal elements.

\section{Discussion}

Finally, we come back to Eq. \ref{eq1}, which
seemingly suggests that the position operator is diagonal in $\bk$.
To reveal the true reason behind this, we first apply the
translational operator $\hat{T}$ to $\br$ to get
$\hat{T}\br=\br+\bR_l$, where $\bR_l$ is the lattice vector of a
crystal. This proves that the position operator is not translationally
invariant, but if we apply it to the commutation \cite{esteve2023}
$[\hr,\hH]$, where $\hH$ is the crystal Hamiltonian operator and has
the periodicity of the crystal, we have
$\hat{T}[\hr,\hH]=[\hr,\hH]$. This shows the commutation itself is
cell-periodic, which explains why Eq. \ref{eq1} is diagonal in
$\bk$. However, this does not prove $\hr$ is translationally invariant
and diagonal in $\bk$. Instead, Eq. \ref{eq1} only means that only
diagonal matrix elements of $\bp$ and $\hr$ are related to each other.
For the rest, they are not, because $\hat{T}[\hr^n,\hH]\ne
[\hr^n,\hH]$, where order $n$ is larger than 1.

\section{Conclusion}

{By directly computing the matrix elements of the
position operator between two Bloch states, we have found a resolution
to a long-standing problem with the position operator.  We show} that
the position operator is a full matrix in $\bk$ and band index $n$ in
general, in contrast to prior theories \cite{blount1962,aversa1995}.
The fundamental reason why it must be so is because the position
operator is not translationally invariant. Whenever the position
operator appears along a particular direction, the translational
symmetry is broken, which introduces a phase factor, instead of a
Kronecker $\delta$.  Our finding is expected to have a significant
impact on a large group of prior studies from 1990s up to now
\cite{aversa1995,hughes1996,sharma2003,souza2004,nastos2005,nastos2006,virk2007,luppi2010,margulis2010,attaccalite2011,bgu2013,taghizadeh2017,hipolito2018,yue2020,thong2021,liebscher2021,gu2022,vlasluk2023,gassner2023,ngo2023,parks2023,tavakol2023,weitz2024},
including some softwares \cite{vasp,reascos2024}, where off-diagonal
matrix elements of the position operator in $\bk$ are ignored. It
represents a paradigmatic shift for future research.  Specifically,
our findings will open the door to future researchers to compute (i)
the polarization directly using the Bloch wavefunctions, without
transforming them to the Wannier wavefunctions \cite{vanderbilt},
fully respecting the definite parity under space inversion, in
contrast to the exponential operator
\cite{resta1998,aligia1999,evangelisti2022}, (ii) optical transitions
within the length gauge, with no difficulty with degenerate or nearly
degenerate bands \cite{foreman2000}, since no derivative is taken, and
(iii) orbital magnetization for a solid
\cite{xiao2005,thonhauser2005,ceresoli2006,shi2007}.  Some potential
applications may be still beyond our current imagination.

\section*{acknowledgments}
We would like to thank Drs. Michael Berry (Bristol),
Raffaele Resta (Trieste) and Juan Jose Esteve-Paredes (Madrid) for
their helpful communications and suggestions.  GPZ was partly
supported by the U.S. Department of Energy under Contract No.
DE-FG02-06ER46304.  MSS was supported by the National Science
Foundation of China under Grant No. 11874189.  We acknowledge that our
work was carried out on Indiana State University's Quantum and
Obsidian Clusters.

$^{*}$guo-ping.zhang@outlook.com.
https://orcid.org/0000-0002-1792-2701

\clearpage

\clearpage
\newpage

\centerline{\bf Supplementary Materials}

This supplementary material contains additional information to support
our conclusion of the main paper.  In Sec. I, we briefly explain why
the direction of the position operator affects its matrix elements
between two Bloch wavefunctions.  Section II is devoted to the
derivation of the $\bk$-diagonal elements of the position operator.
Section III shows that Eq. \ref{deriv1} is not singular even when
$q_1\rightarrow k_1$.  Section IV provides the details of our matrix
elements of the position operator for the nearly free electron model.

\setcounter{equation}{0}

\setcounter{page}{1}

\setcounter{section}{0}

\section{Crystal momentum and  position operator}

The matrix elements of the position operator $\hr$ between two Bloch
states $\psi^*_n(\bk,\br)$ and $\psi_m(\bq,\br)$ depend on 
$\bk$ and $\bq$ as variables, so 
$\bk$ and $\bq$
may take any values that the system allows.
But some combinations of the pairs $\bk$ and $\bq$ yield null matrix
elements for the position operator.
Following our main text, we take  $\br$ 
along the ${\bf a}_1$ direction as an example. 
 We reproduce Eq. \ref{u4} of the main paper here,
\ba &\frac{1}{V}\int_V\psi^*_n(\bk,\br)r_1 \hat{\bf a}_1
\psi_m(\bq,\br)dr_1 dr_2dr_3 \nonumber \\ =&\frac{\hat{\bf
    a}_1}{V}\int_{r_1}\int_{r_2}\int_{r_3}
u^*_n(\bk,\br)u_m(\bq,\br)e^{
i(q_1-k_1)r_1+i(q_2-k_2)r_2+i(q_3-k_3)r_3}r_1 dr_1
dr_2dr_3.  \label{s1} \ea Along ${\bf a}_2$ and ${\bf a}_3$ the
integral is cell-periodic and must yield two Kronecker
$\delta_{q_2,k_2}$ and $\delta_{q_3,k_3}$, Eq. \ref{s1} is zero unless
$q_2=k_2$ and $q_3=k_3$.  But along ${\bf a}_1$ direction, $q_1$ and
$k_1$ may or may not equal to each other, since the integral is no
longer cell-periodic because of the presence of $r_1$.  If $q_1=k_1$,
then we have a $\bk$-diagonal element for the position; otherwise, we
have a $\bk$-off-diagonal element.  In general, there are two terms,
one diagonal and one off-diagonal in $\bk$ in the matrix elements.
This shows that the direction of $\br$ in the real space affects the
matrix elements. If $\br$ points in an arbitrary direction, then there
are three possible combinations as shown in Eq. \ref{main} of the main
text.

\section{Derivation of the $\bk$-diagonal elements of the position
  operator}

 Equation \ref{u4} of the main paper  (see Eq. \ref{s1} above) 
contains four possible cases: (i) diagonal in
 crystal momentum $\bk$ and diagonal in bands, i.e., $k_1=q_1$ and
 $n=m$, (ii) $k_1=q_1$ and $n\ne m$, (iii) $k_1\ne q_1$ and $n=m$ and
 (iv) $k_1\ne q_1$ and $n\ne m$. We first consider (i) and (ii), so
 all the exponential terms in Eq. \ref{u4} are 1. We note in
 Eq. \ref{u4} $V=N_1N_2N_3\Omega$. Without the exponential terms, the
 integration over $r_2$ ($r_3$) is just $N_2$ ($N_3$) times the
 integral over the primitive cell, so Eq. \ref{u4} can be simplified
 to \ba \frac{\hat{\bf a}_1}{V}\iiint u^*_n(\bk,\br)u_m(\bk,\br)r_1
 dr_1 dr_2dr_3 \nonumber \\ =\frac{\hat{\bf a}_1}{N_1\Omega}
 \int_{r_1=0}^{N_1a_1} \int_{r_2=0}^{a_2}\int_{r_3=0}^{a_3}
 u^*_n(\bk,\br)u_m(\bk,\br)r_1 dr_1 dr_2dr_3. \label{u41} \ea Then we
 break the integration over $r_1$ into $N_1$ segments, each of which
 spans a unit cell, i.e.,
 $\int_{r_1=0}^{N_1a_1}=\int_{r_1=0}^{a_1}+\int_{r_1=a_1}^{2a_1}+\cdots+
 \int_{r_1=(N_1-1)a_1}^{N_1a_1}$. Except the first integral, the rest
 of integrals lie outside the primitive unit cell, i.e., $r_1$ beyond
 the primitive cell.  What we will do is to express these remaining
 integrals in terms of the first integral $\int_{r_1=0}^{a_1}$, plus
 an extra term. For the illustrative purpose, we take a generic one as
 an example (the integrations over $r_2$ and $r_3$ are implied). We
 replace the integration over $r_1$ by over $\xi_1=r_1-ja_1$ to obtain
 \ba \int_{r_1=ja_1}^{(j+1)a_1} u^*_n(\bk,\br)u_m(\bk,\br)r_1 dr_1
 =\int_{\xi=0}^{a_1} u^*_n(\bk,\br)u_m(\bk,\br)(\xi_1+ja_1) dr_1
 \nonumber \\ =\int_{\xi_1=0}^{a_1} u^*_n(\bk,\br)u_m(\bk,\br)\xi_1
 d\xi_1+ja_1 \int_{\xi_1=0}^{a_1} u^*_n(\bk,\br)u_m(\bk,\br) d\xi_1,
 \nonumber \ea where the first term is just $\int_{r_1=0}^{a_1}$ and
 the integral in the second term is the same for every cell. { So our
 integral over $r_1$ is now
 \ba
& \int_{r_1=0}^{N_1a_1}u^*_n(\bk,\br)u_m(\bk,\br)r_1 dr_1
\nonumber \\&=N_1\int_{r_1=0}^{a_1}  u^*_n(\bk,\br)u_m(\bk,\br)r_1 dr_1  + \frac{(N_1-1)N_1a_1}{2}
 \int_{r_1=0}^{a_1} u^*_n(\bk,\br)u_m(\bk,\br) dr_1,
 \nonumber
 \ea}
 which is
 plugged into Eq. \ref{u41} to find \ba \frac{\hat{\bf a}_1}{\Omega}
 \int_{r_1=0}^{a_1} \int_{r_2=0}^{a_2}\int_{r_3=0}^{a_3}
 u^*_n(\bk,\br)u_m(\bk,\br)r_1 dr_1 dr_2dr_3\nonumber
 \\ =\frac{\hat{\bf a}_1}{\Omega} \left (\int_\Omega
 u^*_n(\bk,\br)u_m(\bk,\br)r_1 dr_1
 dr_2dr_3+\frac{(N_1-1)a_1}{2}\int_\Omega u^*_n(\bk,\br)u_m(\bk,\br)
 dr_1dr_2dr_3\right ),
 \label{u42}
\ea 
where the first term is the same for
 all the cells along 
$\hat{\bf a}_1$. The second term appears when $r_1$ in Eq. \ref{u4} 
exceeds $a_1$.  If $n=m$, both terms are kept, and they are 
diagonal in $\bk$ and $n$. In this case, 
the matrix elements depend on the  origin of $r_1$, which is why
choosing a different unit cell can change the matrix element
 \cite{vanderbilt}.  \UU{change4} {For example, 
 we consider two 1D $u$ functions,
 $u_1(x)=\frac{1}{\sqrt{\pi}} 
\sin(x+1)$ and  $u_2(x)=\frac{1}{\sqrt{\pi}} 
\sin (x)$, integrated over $[-\pi,\pi]$. $u_1(x)$ is shifted with
 respect to $u_2(x)$ by 1. If we compute the diagonal element of the
 position operator, we obtain $\la u_1|x|u_1\ra=\int_{-\pi}^{\pi}
 x \sin^2(x+1) dx=-\frac{1}{2}\pi\sin(2)$ and  $\la u_2|x|u_2\ra
 =\int_{-\pi}^{\pi} x\sin^2x dx =0$. So the matrix elements are
 different. }
 If $n\ne m$, i.e., diagonal in $\bk$ but
 off-diagonal in $n$, the second term is zero due to the orthogonality
 and the first term is independent of the origin of $r_1$.

\section{Derivation of Eq. \ref{deriv1}}

\newcommand{\qq}{i(q_1-k_1)a_1}

Here we prove Eq. \ref{deriv1} of our main text 
 does not have a singularity even when
$q_1\rightarrow k_1$.
We start from Eq. \ref{deriv1}
\ba
\frac{-iN_1a_1e^{iN_1(q_1-k_1)a_1}}{1-e^{i(q_1-k_1)a_{1}}}
+\frac{(1-e^{iN_1(q_1-k_1)a_1})(ia_1e^{i(q_1-k_1)a_1})}
{(1-e^{i(q_1-k_1)a_{1}})^2} \\
=ia_1\frac{-N_1e^{iN_1(q_1-k_1)a_1}(1-e^{i(q_1-k_1)a_{1}})+
(1-e^{iN_1(q_1-k_1)a_1})e^{i(q_1-k_1)a_1} 
}
{(1-e^{i(q_1-k_1)a_{1}})^2}\\
=ia_1\frac{-N_1e^{iN_1(q_1-k_1)a_1}+
e^{i(q_1-k_1)a_1}+(N_1-1) 
e^{i(N_1+1)(q_1-k_1)a_1})
}
{(1-e^{i(q_1-k_1)a_{1}})^2} \label{der2}
.
\ea
We first examine the denominator by expanding the exponent in terms of
$(q_1-k_1)a_1$  
\ba
(1-e^{i(q_1-k_1)a_{1}})^2=\left [1-1-i(q_1-k_1)a_1-\frac{1}{2!}(i\left
  (q_1-k_1)a_1\right)^2+\cdots\right]^2\\
= [\qq]^2 [1+\frac{1}{2}\qq+\cdots]^2,
\ea
which shows the leading order is $[\qq]^2$. This means that the
numerator in Eq. \ref{der2} must be expanded at least to  $[\qq]^2$.

The numerator  in Eq. \ref{der2}
is
\ba
-N_1e^{iN_1(q_1-k_1)a_1}+e^{i(q_1-k_1)a_1}+(N_1-1)e^{i(N_1+1)(q_1-k_1)a_1}
\nonumber \\
=-N_1(1+iN_1(q_1-k_1)a_1+\frac{(iN_1(q_1-k_1)a_1)^2}{2}+\cdots) \nonumber\\
+1+\qq+\frac{1}{2}\qq^2+\cdots \nonumber\\
+(N_1-1)(1+i(N_1+1)(q_1-k_1)a_1+\frac{(i(N_1+1)(q_1-k_1)a_1)^2}{2}+\cdots,
\ea
whose zeroth order in $\qq$ is $-N_1+1+N_1-1=0$, and whose first order 
is $-iN_1^2(q_1-k_1)a_1+i(q_1-k_1)a_1+i(N_1^2-1)(q_1-k_1)a_1=0$. 
The second order is
\be
\frac{(\qq)^2}{2}(N_1^2-N_1)=-\frac{(q_1-k_1)^2a_1^2}{2}(N_1^2-N_1),
\ee
which is the lowest nonzero term.  Equation \ref{deriv1} in the limit
$q_1\rightarrow k_1$ is given
by
\be
ia_1\frac{-\frac{(q_1-k_1)^2a_1^2}{2}(N_1^2-N_1)}
{[\qq]^2 [1+\frac{1}{2}\qq+\cdots]^2} =\frac{ia_1(N_1^2-N_1)}{2}, \label{der3}
\ee
If we divide Eq. \ref{der3} by $i$, we find
$\frac{a_1(N_1^2-N_1)}{2}$, which matches our above result obtained by
the direct calculation. Here we do not use the BvK boundary condition.

\section{Nearly free-electron model}

We first show some details how to get the first-order correction to
the wavefunction.  We denote the first-order correction to the
wavefunction as $\psi^{(1)}_k$, which is given by \ba
\psi^{(1)}_k=\sum_{q\ne k} \frac{\la \phi_q|\sum_n U_n
  e^{inbx}|\phi_k\ra}{E_k^{(0)} -E_q^{(0)}} |\phi_q\ra =\sum_{q\ne
  k}\frac{\sum_n U_n\delta_{k+nb,q}}{E_k^{(0)} -E_q^{(0)}}
\frac{1}{\sqrt{L}}e^{iqx}\\ =
\frac{1}{\sqrt{L}}\sum_{n\ne0}\frac{U_n}{E_k^{(0)} -E_{k+nb}^{(0)}}
e^{i(nb+k)x} = \frac{1}{\sqrt{L}}\sum_{n\ne0}\frac{U_n
  e^{inbx}}{E_k^{(0)} -E_{k+nb}^{(0)}} e^{ikx}. \ea 
Including the zeroth-order wavefunction, 
we have  our wavefunction up
to the first order correction as \be \psi(k,x)=\frac{{\cal C}}
{\sqrt{L}}\left(1+\sum_{n\ne0}\frac{U_{n}}{
  \frac{\hbar^{2}k^{2}}{2m_e}-\frac{\hbar^{2}(k+nb)^{2}}{2m_e}
}e^{inbx}\right) e^{ikx}
\equiv \frac{{\cal C}}{\sqrt{L}}\left (1+\sum_{n\ne0} W_n(k) e^{inbx}\right)e^{ikx}
\label{sup.eq1}
\ee which is our Eq. \ref{ch6.eq1212c}
 in the main text.  Here ${\cal C}$ is the
normalization constant, $b=\frac{2\pi}{a}$, $U_n$ is the Fourier
coefficient of $U(x)$, and  the dimensionless $W_n$ is  defined through Eq. \ref{sup.eq1}.

The matrix element of the
position operator for $k_1\ne k_2$ consists of four integrals,
\ba
\la \psi(k_1,x)|\hx|\psi(k_2,x)\ra =\frac{|{\cal C}|^2}{L}\int_0^L
\left ( e^{-ik_1x+ik_2x}x + 
\sum_{m\ne0} 
W_m(k_2)e^{imbx+ik_2x-k_1x} x \right . \nonumber \\
\left . +\sum_{n\ne0}W^*_n(k_1)e^{i(k_2-nb-k_1)x}x+\sum_{n\ne0}\sum_{m\ne0}W^*_n(k_1)W_m(k_2)
e^{i(mb+k_2-nb-k_1)} x\right ) dx\nonumber\\
=\frac{|{\cal C}|^2}{i(k_2-k_1)}
+\sum_{m\ne0}\frac{|{\cal C}|^2W_m(k_2)}{i(mb+k_2-k_1)}
+\sum_{n\ne0}\frac{|{\cal C}|^2W_n^*(k_1)}{i(k_2-nb-k_1)}+\sum_{n,m\ne0}\frac{|{\cal
    C}|^2W_n^*(k_1) W_m(k_2)}{i(mb+k_2-nb-k_1),} \nonumber
\ea
which is Eq. 11 of the main paper.


\begin{thebibliography}{99}

\bibitem{dirac}P. A. M. Dirac, {\it The Principles of Quantum
Mechanics},  Oxford Science Publications, Clarendon Press, 4th
ed. (1982).

\bibitem{schiff} L. Schiff, {\it Quantum Mechanics}, McGraw-Hill, New
York, second edition (1955).

\bibitem{messiah} A. Messiah, {\it Quantum Mechanics}, North-Holland,
Amsterdam (1961).

\bibitem{landau} L. Landau and E. Lifshitz, {\it Quantum Mechanics},
Addison-Wesley (1958).

\bibitem{griffiths2018} D. J. Griffiths and D. F. Schroeter, {\it
Introduction to quantum mechanics}, Cambridge University Press, 3rd
edition, page 182 (2018).

\bibitem{sakurai} J. J. Sakurai, {Modern quantum mechanics}, ed. by
S. F. Tuan, Pearson (2005).

\bibitem{ourqm}Guo-ping Zhang, Mingsu Si and Thomas F. George,
{\it Quantum Mechanics}, De Gruyter (2024).


\bibitem{kittel}C. Kittel, {\it Introduction to Solid State Physics},
7th Ed., John Wiley \& Sons, Inc., New York (1996).

\bibitem{blount1962}E. I. Blount, {Formalisms of Band Theory}, Solid
State Phys. {\bf 13}, 305 (1962).

\bibitem{sipe2000}J. E. Sipe and A. I. Shkrebtii, {Second-order
optical response in semiconductors}, Phys. Rev. B {\bf 61}, 5337
(2000).

\bibitem{marzari2012}N. Marzari, A. A. Mostofi, J. R. Yates, I. Souza and
D. Vanderbilt, {Maximally localized Wannier functions: Theory and
applications}, Rev. Mod. Phys. {\bf 84}, 1419 (2012).

\bibitem{bgu2013}B. Gu, N. H. Kwong, and R. Binder, {Relation between
the interband dipole and momentum matrix elements in
semiconductors}, Phys. Rev. B {\bf 87}, 125301 (2013).

\bibitem{butcher}P. N. Butcher and D. Cotter, {\it The Elements of
Nonlinear Optics}, Cambridge University Press, Cambridge (1990).

\bibitem{shen}Y. R. Shen, {\it The Principles of Nonlinear Optics},
John Wiley \& Sons, Inc., Hoboken, New Jersey (2003).

\bibitem{martin1974}R. M. Martin, {Comment on calculation
electric polarization in crystals}, Phys. Rev. B {\bf 9}, 1998
(1974).

\bibitem{asboth2015}J. K. Asboth, L. Oroszlany, and A. Palyi, {\it A
Short Course on Topological Insulators}, Springer (2016).

\bibitem{vanderbilt}D. Vanderbilt, {\it Berry Phases in Electronic
Structure Theory},  Cambridge University Press, Cambridge (2018).

\bibitem{osika2017}E. N. Osika, A. Chacon, L. Ortmann, N. Sarez,
J. A. Perez-Hernandex, B. Szafran, M. F. Ciappina, A. S. Landsman
and M. Lewenstein, {Wannier-Bloch approach to localization in
high-harmonics generation in solids}, Phys. Rev. X {\bf 7}, 021017
(2017).

\bibitem{parks2023}A. M. Parks, J. V. Moloney, and T. Brabec, {Gauge
invariant formulation of the semiconductor Bloch equations},
Phys. Rev. Lett. {\bf 131}, 236902 (2023).

\bibitem{resta1998}R. Resta, {Quantum-mechanical position operator in
extended systems}, Phys. Rev. Lett. {\bf 80}, 1800 (1998).

\bibitem{aligia1999}A. A. Aligia and G. Ortiz, {Quantum mechanical
position operator and localization in extended systems},
Phys. Rev. Lett. {\bf 82}, 2560 (1999).

\bibitem{evangelisti2022} S. Evangelisti, F. Abu-Shoga, C. Angeli,
G. L. Bendazzoli, and J. A. Berger, {Unique one-body position
operator for periodic systems}, Phys. Rev. B {\bf 105}, 235201
(2022).

\bibitem{attaccalite2013}C. Attaccalite and M. Gr\"uning, {Nonlinear
optics from an ab initio approach by means of the dynamical Berry
phase: Application to second- and third-harmonic generation in
semiconductors}, Phys. Rev. B {\bf 88}, 235113  (2013).

\bibitem{draxl2006}C. Ambrosch-Draxl and O. Sofo, {Linear optical
properties of solids within the full-potential linearized augmented
planewave method}, Comp. Phys. Comm. {\bf 175}, 1 (2006).

\bibitem{cazzaniga2010}M. Cazzaniga, L. Caramella, N. Manini and
G. Onida, {Ab initio intraband contributions to the optical
properties of metals}, Phys. Rev. B {\bf 82}, 035104 (2010).

\bibitem{souza2004}I. Souza, J. Iniguez and D. Vanderbilt, {Dynamics
of Berry-phase polarization in time-dependent electric fields},
Phys. Rev. B {\bf 69}, 085106 (2004).

\bibitem{hughes1996} J. L. P. Hughes and J. E. Sipe, {Calculation of
second-order optical response in semiconductors}, Phys. Rev. B {\bf
53}, 10751 (1996).

\bibitem{prb09} G. P. Zhang, Y. H. Bai, and T. F. George, {Energy- and
crystal momentum-resolved study of laser-induced femtosecond
magnetism}, Phys. Rev. B {\bf 80}, 214415 (2009).

\bibitem{esteve2023} J. J. Esteve-Paredes and J. J. Palacios, {A
comprehensive study of the velocity, momentum and position
matrix elements for Bloch states: Application to a local
orbital basis},  SciPost Phys. Core {\bf 6}, 002 (2023).

\bibitem{aspnes1972} D. E. Aspnes, {Energy-band theory of the
second-order nonlinear optical susceptibility of crystals of
zinc-blend symmetry}, Phys. Rev. B {\bf 6}, 4648 (1972).

\bibitem{ghahramani1991} E. Ghahramani, D. J. Moss, and J. E. Sipe,
{Full-band-structure calculation of second-harmonic generation in
odd-period strained (Si)$_n$/(Ge)$_n$ superlattices}, Phys. Rev. B
{\bf 43}, 8990 (1991).

\bibitem{leitsmann2005} R. Leitsmann, W. G. Schmidt, P. H. Hahn, and
F. Bechstedt, {Second-harmonic polarizability including
electron-hole attraction from band-structure theory},
Phys. Rev. B {\bf 71}, 195209 (2005).

\bibitem{allen} P. B. Allen, {Electron Transport}, {\it Conceptual
foundations of Materials: A standard model for Ground- and
Excited-State Properties}, Ed. S. Louie and M. Cohen (2006). Page
165.

\bibitem{sharma2003} S. Sharma, J. K. Dewhurst, and C. Ambrosch-Draxl,
{Linear and second-order optical response of III-V monolayer
superlattices}, Phys. Rev. B {\bf 67}, 165332 (2003).

\bibitem{karplus1952} R. Karplus and J. M. Luttinger, {Hall effect in
ferromagnetics}, Phys. Rev. {\bf 95}, 1154 (1952).

\bibitem{callaway} J. Callaway, {\it Quantum Theory of the Solid
State}, Academic Press, Inc., New York (1974).

\bibitem{lax} M. Lax, {\it Symmetry Principles in Solid State and Molecular
Physics}, J. Wiley and Sons, Inc., New York (1974).

\bibitem{matsyshya2019} O. Matsyshyn and I. Sodemann, {Nonlinear Hall
acceleration and the quantum rectification sum rule},
Phys. Rev. Lett. {\bf 123}, 246602 (2019).

\bibitem{aversa1994} C. Aversa, J. E. Sipe, M. Sheik-Bahae and
E. W. Van Stryland, {Third-order optical nonlinearities in
semiconductors: The two-band model}, Phys. Rev. B {\bf 50}, 18073
(1994).

\bibitem{aversa1995} C. Aversa and J. E. Sipe, {Nonlinear optical
susceptibilities of semiconductors: Results with a length-gauge
analysis}, Phys. Rev. B   {\bf 52}, 14636 (1995).

\bibitem{virk2007} K. S. Virk and J. E. Sipe, {Semiconductor optics in
length gauge: A general numerical approach}, Phys. Rev. B {\bf 76},
035213 (2007).





\bibitem{vasp} G. Kresse and J. Furthm\"{u}ller,
Comput. Mater. Sci. \textbf{6}, 15 (1996).

\bibitem{foreman2000} B. A. Foreman, {Theory of the effective
Hamiltonian for degenerate bands in an electric field},
  J. Phys.: Condens. Matter {\bf 12}, R435 (2000).


\bibitem{sm} Supplementary materials contain additional details of
  the $\bk$-diagonal elements of the position operator, no divergence
  even when $q_1\rightarrow k_1$,  our matrix elements of the
  position operator for the nearly free electron model.
  



\bibitem{zak1982} J. Zak, {Band center-A conserved quantity in
  solids}, Phys. Rev. Lett. {\bf 48}, 359 (1982).


\bibitem{shirley1996}E. L. Shirley, {Optimal basis sets for detailed
Brillouin-zone integrations}, Phys. Rev. B {\bf 54}, 16464 (1996).

\bibitem{prendergast2009}D. Prendergast and S. G. Louie,
{Bloch-state-based interpolation: An efficient generalization of the
Shirley approach to interpolating electronic structure},
Phys. Rev. B {\bf 80}, 235126 (2009).




\bibitem{zak1989}J. Zak, {Berry's phase for energy bands in solids},
  Phys. Rev. Lett. {\bf 62}, 2747 (1989).




\bibitem{prb00} G. P. Zhang, {Origin of the 0.89 eV peak in
$\chi^{(3)}(-3\omega;\omega,\omega,\omega)$ of polyacetylene:
Electron correlation effects}, Phys. Rev. B {\bf 61}, 4377 (2000).

\bibitem{nastos2005} F. Nastos, B. Olejnik, K. Schwarz, and
J. E. Sipe, {Scissors implementation within length-gauge formulations
of the frequency-dependent nonlinear optical response of
semiconductors}, Phys. Rev. B {\bf 72}, 045223 (2005).

\bibitem{thong2021}L. H. Thong, C. Ngo, H. T. Duc, X. Song,
and T. Meier, {Microscopic analysis of high harmonic
generation in semiconductors with degenerate bands},
Phys. Rev. B {\bf 103}, 085201 (2021).


\bibitem{nastos2006} F. Nastos and J. E. Sipe, {Optical rectification
and shift currents in GaAs and GaP response: Below and above the
band gap}, Phys. Rev. B {\bf 74}, 035201 (2006).

\bibitem{luppi2010} E. Luppi, H. Hubener, and V. Veniard, {Ab initio
second-order nonlinear optics in solids: Second-harmonic generation
spectroscopy from time-dependent density-functional theory},
Phys. Rev. B {\bf 82}, 235201 (2010).

\bibitem{margulis2010} Vl. A. Margulis, E. E. Muryumin, and
E. A. Gaiduk, {Second-order nonlinear optical response of zigzag BN
single-walled nanotubes}, Phys. Rev. B {\bf 82}, 235426 (2010).

\bibitem{attaccalite2011}C. Attaccalite, M. Gr\"uning, and A. Marini,
{Real-time approach to the optical properties of solids and
nanostructures: Time-dependent Bethe-Salpeter equation},
Phys. Rev. B {\bf 84}, 245110 (2011).

\bibitem{taghizadeh2017} A. Taghizadeh, F. Hipolito, and
T. G. Pedersen, {Linear and nonlinear optical response of crystals
using length and velocity gauges: Effect of basis truncation},
Phys. Rev. B {\bf 96}, 195413 (2017).

\bibitem{hipolito2018} F. Hipolito, A. Taghizadeh, and T. G. Pedersen,
{Nonlinear optical response of doped monolayer and bilayer graphene:
Length gauge tight-binding model}, Phys. Rev. B {\bf 98}, 205420
(2018).

\bibitem{yue2020} L. Yue and M. B. Gaarde, {Structure gauges and laser
gauges for the semiconductor Bloch equations in high-order harmonic
generation in solids}, Phys. Rev. A {\bf 101}, 053411 (2020).

\bibitem{liebscher2021} S. C. Liebscher, M. K. Hagen, J. Hader, J.
V. Moloney, and S. W. Koch, {Microscopic theory for the incoherent
resonant and coherent off-resonant optical response of tellurium},
Phys. Rev. B {\bf 104}, 165201 (2021).

\bibitem{gu2022} J. Gu and M. Kolesik, {Full-Brillouin-zone
calculation of high-order harmonic generation from solid-state
media}, Phys. Rev. A {\bf 106}, 063516 (2022).

\bibitem{vlasluk2023} E. Vlasiuk, V. K. Kozin, J. Klinovaja, D. Loss,
I. V. Iorsh, and I. V. Tokatly, {Cavity-induced charge transfer in
periodic systems: Length-gauge formalism}, Phys. Rev. B {\bf 108},
085410 (2023).

\bibitem{gassner2023} S. Gassner and E. J. Mele, {Regularized
lattice theory for spatially dispersive nonlinear optical
conductivities}, Phys. Rev. B {\bf 108}, 085403 (2023)

\bibitem{ngo2023} C. Ngo, S. Priyadarshi, H. T. Duc, M. Bieler, and
T. Meier, {Excitonic anomalous currents in semiconductor quantum
wells,} Phys. Rev. B {\bf 108}, 165302 (2023).

\bibitem{tavakol2023} O. Tavakol and Y. B. Kim, {Nonlinear optical
responses in nodal line semimetals}, Phys. Rev. B {\bf 107}, 035114
(2023).

\bibitem{weitz2024} T. Weitz, C. Heide, and P. Hommelhoff,
{Strong-field Bloch electron interferometry for band-structure
retrieval}, Phys. Rev. Lett. {\bf 132}, 206901 (2024).

\bibitem{reascos2024}L. Reascos, F. Carneiro, A. Pereira, N. F.
Castro and R. M. Ribeiro, Berry: A code for the differentiation of
Bloch wavefunctions from DFT calculations, Comput. Phys.
Commun. {\bf 295}, 108972 (2024).

\bibitem{xiao2005}D. Xiao, J. Shi, and Q. Niu, {Berry phase correction
to electron density of states in solids}, Phys. Rev. Lett. 95, 137204
(2005).

\bibitem{thonhauser2005} T. Thonhauser, D. Ceresoli, D. Vanderbilt,
and R. Resta, {Orbital magnetization in periodic insulators},
Phys. Rev. Lett. {\bf 95}, 137205 (2005).

\bibitem{ceresoli2006} D. Ceresoli, T. Thonhauser, D. Vanderbilt, and
R. Resta, {Orbital magnetization in crystalline solids: Multi-band
insulators, Chern insulators, and metals}, Phys. Rev. B {\bf 74},
024408 (2006).

\bibitem{shi2007}J. Shi, G. Vignale, D. Xiao, and Q. Niu, {Quantum
theory of orbital magnetization and its generalization to
interacting systems}, Phys. Rev. Lett. {\bf 99}, 197202 (2007).

\end{thebibliography}
\end{document}